\documentclass{article}

\usepackage{amsfonts,xr,graphicx,amsmath}

\begin{document}
\title{{On stationary solutions of two-dimensional Euler Equation} }
 \author{{Nikolai Nadirashvili\thanks{LATP, CMI, 39, rue F. Joliot-Curie, 13453
Marseille  FRANCE, nicolas@cmi.univ-mrs.fr} }}

\date{}
\maketitle

\def\C{\mathbb{C}}
\def\S{\mathbb{S}}
\def\Z{\mathbb{Z}}
\def\R{\mathbb{R}}
\def\N{\mathbb{N}}
\def\H{\mathbb{H}}
\def\tilde{\widetilde}
\def\epsilon{\varepsilon}

\def\n{\hfill\break} \def\al{\alpha} \def\be{\beta} \def\ga{\gamma} \def\Ga{\Gamma}
\def\om{\omega} \def\Om{\Omega} \def\ka{\kappa} \def\lm{\lambda} \def\Lm{\Lambda}
\def\dd{\delta} \def\Dl{\Delta} \def\vph{\varphi} \def\vep{\varepsilon} \def\th{\theta}
\def\Th{\Theta} \def\vth{\vartheta} \def\sg{\sigma} \def\Sg{\Sigma}
\def\bendproof{$\hfill \blacksquare$} \def\wendproof{$\hfill \square$}
\def\holim{\mathop{\rm holim}} \def\span{{\rm span}} \def\mod{{\rm mod}}
\def\rank{{\rm rank}} \def\bsl{{\backslash}}
\def\il{\int\limits} \def\pt{{\partial}} \def\lra{{\longrightarrow}}
\def\pa{\partial } 
\def\ra{\rightarrow }
\def\sm{\setminus }
\def\ss{\subset }
\def\ee{\epsilon }

 {\em Abstract.} We study the geometry of streamlines and stability properties for steady state solutions of the Euler equations for ideal fluid.

\section{Introduction}
\bigskip

Let $\Omega \subset \R^2$ be a bounded domain with a smooth boundary. Let
$v(x,t)= (v_1,v_2)$, $x\in \Omega , t\in \R$ be a solution of the Euler equation for an ideal fluid: 

\begin{equation}\label{1} \left\{
\begin{array}{l l}
\pa v/\pa t +v \nabla v = -\nabla p, &\mbox{in $\Omega\times \R $} \\
div\ v=0 &\mbox{in  $ \Omega \times \R$} \\
\end{array} \right. \end{equation}
 Together with the boundary condition,

\begin{equation}\label{2} 
\begin{array}{l l}
(n,v)=0 &\mbox{on  $\pa \Omega  ,$}
 \end{array}
 \end{equation}
 the equation \eqref{1}, \eqref{2}  defines in spaces $C^{k, a} $, $k=1,2,..., 0<a<1$ an evolution operator,  $ e^t: v(x,0)\lra v(x,t)$,  i.e., for any initial data $v_0\in C^{k, a}(\Omega) $,
$$v(x, 0)= v_0(x)   $$ 
there exists a unique
solutions $v(t,x) $ of (1) defined for all $t\in \R $ such that for all $t\in \R $ $v\in C^{k, a}(\Omega)  $, see [L]. For  analytical  $v_0$ the solution $v$ remains analytic for  all times $t$,
[BBZ], [AM].

Vector field $v(x,t)$ defines a flow $g(x,t)$ on $\Omega $,
$$g:\Omega \ra \Omega , $$
$g$ is one-parametric group of area preserving  diffeomorphisms of $\Omega $. 

Let $x_0\in \Omega $. The curve $\gamma (t)\in \Omega$, 
$$\gamma :t\in\R\ra g(x_0,t)\in\Omega $$
called the streamline of a material particle $x_0$ of the fluid.

Let $\omega =curl\,  v$ be the vorticity of $v$. Then the equation (1) is equivalent to 
Euler-Helmholtz equation for the vertex, [AK],

$$ \omega_t+\omega_v=0.$$

The stationary (or steady) solution are solutions independent on $t$. Therefore the stationary Euler equation is

\begin{equation}\label{3}\left\{
\begin{array}{l l}
v \nabla v +\nabla p =0, &\mbox{in $\Omega\times \R $} \\
div\ v=0 &\mbox{in  $ \Omega \times \R$} \\
\end{array} \right. \end{equation}

In this paper we are concerned wih the structure of steady solutions and the behavior of the flow
$e^t$ in a neighborhood of it. Let $v$ be a solution of \eqref{3}. Streamlines of $v$ are the trajectories of
the corresponding fluid motion, i.e., the integral curves of the vector field $v$. 

\smallskip

{\bf Theorem 1.1.} {\it Let $v\in C^1(\Omega )$ be a steady solution of the Euler equation, $c>|v|>c^{-1}>0$. Then
the streamlines of $v$ are smooth ($C^{\infty } $) curves.  Moreover for any streamline $\gamma $, $C^k$-norms of $\gamma$ at
$x\in \Omega$ depend on the $L_1$-norm  of vorticity $\omega $,  on the constant $c$ and the distant of $x$ to the boundary  of $\Omega $.

If additionally $v\in C^{3,a}, a>0$, then the streamlines are real-analytic. }

\medskip

Theorem 1.1 gives a bound to the acceleration of individual material particles of the flow. In a sense, it explains
a visible boundness of curvature  of  streamlines,  which one can observe
in a lot of experimental pictures of the flow. 

Of course,  in general $v $ is not a smooth vector field on $\Omega $.  The phenomenon  of  a higher regularity of  streamlines  than the regularity of  the solution itself has attracted a lot of attention.
The first underlying ideas to it were  suggested by Lichtenstein [Li]. Another approach to the problem is
connected with the observation of Arnold, [A1]:   flows generated by solutions of the Euler equation \eqref{1}, \eqref{2} can be regarded as geodesics on the group of area preserving diffeomorphisms of
$\Omega $. More general, let $M$ be a smooth $n$-dimensional Riemannian manifold  with a smooth boundary $\partial M$. Denote by $S\, Diff (M)$ the Lie group 
of volume preserving diffeomorphisms of $M$. The tangent space $TM$ are divergent free vector
fields on $M$ tangent to $\partial M$. The scalar product on $TM$ defines a weak right-invariant
metric $g$ on $S\, Diff (M)$, [A1]. 

The geometry of volume preserving diffeomorphisms of a finite
smoothness was studied by Ebin and  Marsden, [EM]. Denote by $D^{1,a}$ the group of volume preserving diffeomorphisms of $M$ which are in $C^{1,a}, a>0$. Notice, the metric $g$ is not complete on $D^{1,a}$.
By the observation of Ebin and Marsden, [EM], metric $g$ defines a {\it smooth }
connection on $D^{1,a}$, and the geodesic exponential map on $(D^{1,a}, g)$,  where it is defined, is smooth, [EM], Theorem 9.1. This  does not imply  that  individual streamlines of the Euler equation are smoothly immersed curves into $M$, but rather the smoothness of the flow in an average sense. We notice that by  the result of 
Milnor there are no analytical structures on $D^{1,a}$, [M], Lemma 9.1, and hence one can not
directly generalize the results of [EM] into an analytic setting. 

Smoothness of the individual streamlines of the Euler equation in $\R^n$ was proved by Chemin, [C],
for the initial data  $v_0\in C^{1,a}, a>0$, so that $C^k$-norms of the streamlines depend on
$C^{1,a}$-norm of  $v_0$. In  Chemin's result the smoothness of the initial
data $v_0$ can not be taken lower than $C^{1,a}, a>0$,  hence the result provides no
bounds for the acceleration of  flow's particles, or for the curvature of streamlines. 

Theorem 1.1 has a local character, we do not assume any boundary condition on $\pa\Omega$. That also distinguish Theorem 1.1 from the previous results.
The proof of the theorem is based on a detail analysis of the elliptic equation for the streamfunction  of the flow.

\smallskip

As a corollary of Theorem 1.1 we
show in Section 2 that the continuity (boundness) of the vorticity implies the continuity (correspondingly,
boundness)  of the first derivatives of the flow $v$. By Yudovich's theorem, [Y],  the 
dynamics $e^t$ of \eqref{1}, \eqref{2} is well defined on the space of divergence free
vector fields $v$ with $\omega \in L_{\infty}$. Thus the last remark means that the Yudovich's
space for the steady flows coincide with the space of Lipschtitz, divergent free vector fields.

From Theorem 1.1 it follows that any individual streamline of a steady flow is defined uniquely by
its any small segment. One can see easily that there is no unique continuation property for the continuation of  
$v$ from subdomain of  $\Omega $ on the whole domain.   

Consider the steady state Euler-Helmholtz equation,
\begin{equation}\label{4} \omega_v=0. \end{equation}
If we  write equation \eqref{4} in the form of first order system we see that the characteristics of \eqref{4} coincide
with the streamlines of the flow $v$. We show that the uniqueness of non-characteristically Cauchy
problem for \eqref{4} requires very low smoothness of the solutions.

{\bf Theorem 1.2.} {\it  Let $v_1,v_2 \in C^1(\bar \Omega )$ be  steady solutions of the Euler equation. Assume that  $\gamma \in C^1$ be an arc on  $\pa \Omega$, and  $v_i$ flow inside $\Omega$ over $\gamma$, i.e., $(v_i, n)>0,\, i=1,2$, where $n$ be the inner normal to $\pa  \Omega $. Assume $v_1=v_2,\, \nabla v_1=\nabla v_2 $ on $\gamma$.  Then streamlines of the flows $v_1$ and $v_2 $  starting from  same points of $\gamma $ coincide. Moreover,  the flows   coincide on the union of these streamlines.}

\smallskip

As a  complement to Theorem 1.1 in the following   theorem we study the structure of
the steady flow in a neighborhood of stagnation points (critical set) of the flow.

{\bf Theorem 1.3.} {\it  Let $v\in C^{1,a}$ be a steady flow defined in $\Omega $. Assume that $0\in \Omega$ is
an isolated critical point of $v$, $v(0)=0$. Then  in a suitable orthonormal coordinates $x_1,x_2$  in a neighborhood of $0$, $v$  has one of the  following expansions

(i) $v=(ax_2,bx_1)+o(|x|)$,  $a,b\neq 0$,

(ii) $v=(\Im (a z^n), \Re (a z^n))+o(|z|^n)$,  $ z=x_1+ix_2\in \C,\, n\in \N, \, n\geq 2,\, a \neq 0$.

(iii) $v=(ax_2+o(x_2), \Re (\alpha z^n )+ o(|x|^n))$,   $\alpha , z=x_1+ix_2\in \C,\, n\geq 2,\,  a, \alpha \neq 0.$

Let $G\ss \ss \Omega$, be a  domain such that  $v=0$ on $G$  and in $\Omega \sm \bar G$ there are no stagnation points of $v$.Then 
there is a neighborhood of $G$ consisting of closed streamlines encircling $G$. }

\smallskip

One can easily see that  in general the stagnation set of a steady flow is not necessarily  discrete. For instance, for  rotationally symmetric steady flows the stagnation set can be a disk. As one can immediately see for the rotationally symmetric flow the domain of stagnation is always encircled
by closed streamlines.  The existence and the structure of Cantor  type stagnation sets 
remains unclear.

Let $v$ be a steady state solution of \eqref{2}. Denote by $u$
the stream function of the flow $v$: $v= (\pa u/\pa y, -\pa u/\pa x)$. Then we have
$$\Delta u=\omega ,$$
 and hence the gradients $\nabla u$ and $\nabla \Delta u$ are parallel.
 A steady solution $v\in C^{1,a}$ called stable in the sense of Arnold, if the stream function $u$ satisfies the
 inequalities:
\begin{equation}\label{5}c< \nabla u /\nabla \Delta u < C,\end{equation}
where $c,C$ are positive constants.

\smallskip

{\bf Theorem 1.4.} {\it Let $\Omega \subset \R^2$ be a bounded convex domain. Let $v$ be a stable in the sense of Arnold  steady state
solution of \eqref{2}, \eqref{3}. Then $v$ has a single critical point.}

\smallskip

We guess that the streamlines in Theorem 1.4 are convex curves, however, for general stationary solutions of \eqref{2}, \eqref{3} in a convex domain there is no convex property for streamlines, [HNY].

 By Arnold's theorem, [A2], [AK],   if $v\in C^{1,a}$ is a steady flow stable in the 
 sense of Arnold then $v$ is a (Lyapunov) stable solution of \eqref{1} with respect to the norm $W^{1,2}(\Omega )$, i.e., for any $\delta >0$ there is an $\epsilon >0$ such that any solution of \eqref{1}  in $C^{1,a}$ which is at   $t=0$  in the $\epsilon $-neighborhood of $v$ in the space $W^{1,2}(\Omega )$ will never leave the $\delta $-neighborhood of $v$. As a consequence $v$ is also stable with respect to  
 the norm $W^{1,p}(\Omega )$ for any $2\leq p < \infty $. Unfortunately in the spaces $W^{1,p}(\Omega )$ dynamics of the Euler equation is unknown. 
It is interesting to understand the stability properties of steady solutions of the Euler equation in
different functional spaces, especially in the spaces $C^{k,a}$ where dynamics of \eqref{1}, \eqref{2} is defined.
The following theorem shows that  Arnold's stable solutions are extremely unstable in $C^{2,a}$.

\smallskip

{\bf Theorem 1.5.} {\it Let $\tilde v\in C^{2,a}(A)$ be a steady radially symmetric solution of the Euler equation defined in the annulus
$A=\{ 1<r<2 \},\, r=|x|$, such that the vorticity $\tilde \omega $ satisfies:  $\tilde \omega >0,\,  \tilde \omega_r >0$ (and therefore  $\tilde v$ stable in the sense of Arnold).  Then there exists a neighborhood $G\subset C^{2,a}(A)$ of  $\tilde v$  such that for any $v_0\in G$ the trajectory $v(\cdot , t)$ in $C^{2,a}(A)$ defined by \eqref{1} is either a stationary solution of the Euler equation, or  there is  $t_0>0$ such that  $v(\cdot , t_0)$ is not in
 $G$. }
 
 \medskip
 
 In particular from Theorem 1.5 it follows that in a neighborhood of $\tilde v$ there are no
 periodic or quasi-periodic solutions of the Euler equation.
 
 In [N] we proved that in any $C^{1,a}$-neighborhood of $\tilde v $ there is $v_0$ such that
 $v$ is a wandering trajectory in $C^{1,a}(A)$. Shnirelman proved, [S], that a typical trajectory of the Euler equation is wandering. Koch shown, [K], that stable in $C^{1,a}$ steady state solution
of the Euler equation generates a periodic flow. These results make natural the following conjecture:

{\bf Conjecture.} {\it There are no stable in $C^{1,a}$ stationary solutions of the Euler equation. }

\medskip

Let $v$ be a steady flow \eqref{2}, \eqref{3} in $\Omega $. Then $v$ is an
extremal of the kinetic energy of $v$,
$$E(v) ={1\over 2}\int_{\Omega }v^2dx,$$
under certain constraints.

 Denote by $S\, Diff\, (\Omega )$ the group of area preserving diffeomorphisms of the domain $\Omega $. We say that two divergence free vector fields $v_1$ and $v_2$ are isovorticed  if their vorticity functions  are the same up to  $S\, Diff\, (\Omega )$
 changes of variable $x$, i.e., if we denote by $S(\omega )$ the orbit of $\omega $ under the action of
 $S\, Diff\, (\Omega )$, we say  $v_1$ and $v_2$ are isovorticed if $\omega_1\in S(\omega_2)$. 
 
 Kelvin noticed that the steady flow $v$ is an extremal of the kinetic energy  $E(v)$ over the set
 of isovorticed with $v$ divergence free vector fields.
 
 Thus  it is  natural to consider the following variational  problem (K): for a given function $h$  on $\Omega $, $h= $const  on $\pa \Omega $, find
 minimizers of the kinetic energy over divergence free vector fields $v$ with  vorticities in $S(h)$.

 There are certain obstructions  for the existence of a smooth minimizer $v$ even if $h$ is smooth,
 see Section 5. Thus, one can try to look for minimizers in the strong closure $\bar S(h)$, which coincide with the set of
 rearrangements of the function $h$, see, e.g., [AP].  Burton proved, [B2], that for any  positive $h\in L_p(\Omega )$,
 $1<p< \infty $, there exists a minimizer (and also a maximizer) of the variational problem (K), 
 $\omega \in \bar S(h)$, and $v,\, curl v=\omega$, is a steady state solution of \eqref{2}, \eqref{3}. Thus
 $v\in W^{1,p}$. Questions remain for smooth $h$. First, for a general smooth $h$ one can expect a
 better than  $v\in W^{1,p}$  smoothness  solutions of (K) . Secondary,
 under some natural assumptions on $h$ can one  expect the existences of a smooth minimizer,
 i.e., a minimizer in $S(h)$. Regarding the second question we prove the following theorem.

  \smallskip

{\bf Theorem 1.6.} {\it  Let $\Omega \ss \R^2$ be a bounded convex domain Let $h$ be a smooth function on $\Omega$ and $h= $const  on $\pa \Omega $. Assume that $h<0$ and has a single critical point in $\Omega $. Then the global minimizer of the variational problem (K) is a smooth in $\Omega \sm \{ s\}$
steady flow $v$ , where $\{s\}$ is a single critical point of $v$ in $\Omega$. Moreover $v$ is an Arnold stable steady state flow in $\Omega $.  }

{\bf Remark.} Notice that if $h$ has a degenerate critical point then $v$ might be not smooth
($C^{\infty }$) in $\Omega $.

It is easy to show that without assumptions on convexity of $\Omega $ or on single critical point
of $h$ the minimizer $v$ might be not smooth. However, we guess that even without these
assumptions the minimizer $v$ will be in $C^{1,a}$.

\medskip

We show that Arnold steady state solution has vorticity of a constant sign. As a consequence we prove that in a convex domain the minimizers of variational
problem (K) are exactly Arnold's stable steady state solutions.

  \smallskip
  
  {\bf Acknowledgements. } The author would like to thank S. Kuksin for very useful discussions.

\section{Level sets of semilinear elliptic equations }

Let $\Omega \subset  \R^n $ be a bounded domain. Let $u\in W^{2,p}$, $p>n$, be a solution of the equation
\begin{equation}\label{6}
\begin{array}{l l}
\Delta u =f(u)  \,\,\, in \,\,\,\Omega \\
\end{array}  \end{equation} 
where $f\in L_p$.

In this section we study the regularity of  level sets of the solution $u$. 

{\bf Theorem 2.1.} {\it Let $x_0\in \Omega $ . Assume that $|u|<C$, $|\nabla u| >c> 0$ in $\Omega $.  Then the level
set $\Gamma = \{ x: u(x)=u(x_0) \}$, $x_0\in \Omega $, is a smooth, $C^{\infty }$, surface and the normal derivative,
$\pa u/\pa \nu $, is a smooth function on $\Gamma$.  The
$C^k$-norms of $\Gamma $ and of $\pa u/\pa \nu $ on $\Gamma $  at $x_0$ are bounded by $C$-norm of $u$ in $\Omega$, constant $c$,  $L_1$-norm of $f$ and the distant of $x_0$
to the boundary $\pa \Omega $.

 If $\pa \Omega \in C^{k,a} $, $x_0\in \pa \Omega$ and in a neighborhood of $x_0$ $u=0$ on $\pa \Omega )  $ then the level sets of
 $u$ in the neighborhood of $x_0$ have uniformly bounded $C^{k,a} $-norm }.
 
 \smallskip
 
 Theorem 2.1 shows a higher smoothness of level sets of   \eqref{6} than it 
follows from the Schauder estimates. It is interesting to compare  the theorem with the result of [HON] where we proved an additional regularity for nodal sets of solutions of a linear Schr\"odinger
 equation.
 
  \smallskip
 
 {\bf Corollary 2.2.} {\it In the assumption of Theorem 2.1 let $f\in C(\R)$ ($f\in L_{\infty}$) Then $u\in C^2$ (correspondingly, $D^2u\in L_{\infty }$) in a neighborhood of $x_0$.}
 
 \smallskip

{\bf Theorem 2.3.} {\it Assume that $n=2$ and $f\in C^{2,a}$-function. Let $x_0\in \Omega $  and let $\nabla u(x_0) \neq 0$ in $\Omega $. Then the level
set $\Gamma = \{ x: u(x)=u(x_0) \}$ is a real-analytic curve nad the normal derivative $\pa u/\pa \nu $ is a real analytic function on $\Gamma $. }

The similar result with a silmilar proof holds in dimension $n$.

Let $u_1, u_2$ be two solutions of \eqref{6} defined in $\Omega $. We prove unique continuation results
for the difference $u_1-u_2$. For sufficiently regular function $f$ the results are well known.

{\bf Theorem 2.4.}  {\it Assume that $n=2$, $\pa \Omega \in C^1$   and $f\in C(\bar \Omega )$,
$\gamma $ be an arc of $\pa \Omega $. Let $u_1, u_2$ be two solution of \eqref{6} and $u=u_1-u_2$. Assume that $u=\nabla u =0$ on $\gamma $ and $\nabla u_1$ is not vanishing on
$\gamma $. Then $u_1\equiv u_2$ in $\Omega$. }

\smallskip

Let $H(x)$ be the fundamental solution of Laplace's equation in $\R^n$.

{\bf Proposition 2.5. }  {\it Let $\Gamma \subset \R^n$ be a bounded surface $\Gamma \in C^{k,a}$, let $\mu \in C^{k-1, a}(\Gamma)$.
Let $l=l_{\Gamma}$ be the single layer potential of $\Gamma$ with the density $\mu $:
$$l[\Gamma , \mu ](x)= \int_{y\in \Gamma}\mu (y) H(x-y)ds.$$
Then the  $C^{k+1,a}$ norms of $l$ are finite  in $\R^n\sm \Gamma$ and bounded by the norms of 
$\Gamma $ and $\mu $ in $C^{k,a}$ and $C^{k-1,a}$ correspondingly. }

\smallskip

Proposition 2.5 is common knowledge. In such generality one can find a proof of the proposition in [W]. For $C^{1,a}$-surfaces see, e.g., [G], [Mi]. Compare also with  general results for the heat kernel in [Ka].   

The single layer potential has a jump of normal derivative over the surface $\Gamma $. Thus
on the whole space the single layer potential has only regularity 
$$l\in C^{0,1}(\R^n).$$

{\bf Proof of Theorem 2.1.} Let $x_0\in \Omega $ and $\nabla u(x_0)\neq 0$. Let $u(x_0)= t_0$.
Denote by $G_t$ the level surface
$$G_t= \{ x\in  \Omega : u(x)=t\}.$$
We will prove that in the neighborhood of the point $(x_0,t_0)$ the surfaces $G_t$ are smooth. From the equation \eqref{6} it follows that $u\in C^{1,a}$ and hence outside the critical points of the function
$u$ the level surfaces of $u$ are in $ C^{1,a}$.

We will denote by $\cong $ equality between functions up to a smooth
function. Thus if $G(x,y)$ be
the Green's function of the Dirichlet problem in $\Omega $ then 
$$G(x,y)\cong P(x-y) $$ for $x$ in a neighborhood of the point $x_0$. The consideration  in the proof below are local, in  the neighborhood of the point $x$.

We will prove by induction over $k=1,2,...$ that 
\begin{equation}\label{n1}G_t\in C^{k,a}\end{equation}
 and the normal derivatives
\begin{equation}\label{n2}\pa u/ \pa \nu \in C^{k-1,a}(G_t).\end{equation}
 
 We know \eqref{n1} and \eqref{n2}  hold for $k=1$. Assume \eqref{n1}  and \eqref{n2}  hold for $k\in \N $. We prove the implications for $k+1$.

Let $t_1$ be sufficiently close to $t_0$. Without loss we will assume that $u>0$ in $\Omega$. Then we have 
\begin{equation}\label{7}u\cong \int_0^{\infty }f(t)l[G_t(\pa u/\pa \nu )^{-1}]dt.\end{equation}
We break the last integral into the sum of three integrals:
$$u\cong \int_0^{t_1-\epsilon}f(t)l[G_t(\pa u/\pa \nu )^{-1}]dt+\int_{t_1-\epsilon }^{t_1+\epsilon}f(t)l[G_t(\pa u/\pa \nu )^{-1}]dt+\int_{t_1+\epsilon }^{\infty }f(t)l[G_t(\pa u/\pa \nu )^{-1}]dt.$$
Denote 
$$u_1=\int_0^{t_1-\epsilon}f(t)l[G_t(\pa u/\pa \nu )^{-1}]dt+\int_{t_1+\epsilon }^{\infty }f(t)l[G_t(\pa u/\pa \nu )^{-1}]dt,$$
$$u_2=\int_{t_1-\epsilon }^{t_1+\epsilon}f(t)l[G_t(\pa u/\pa \nu )^{-1}]dt.$$

Define vector functions $h_{\ee}, q_{\ee}$ on $G_{t_1}$ taking the restrictions:
$$h_{\ee }=\nabla u_1\;\;\; on \;\;\; G_{t_1},$$
$$q_{\ee }=\nabla u_2\;\;\; on\;\;\;  G_{t_1}.$$

The norm $||h_{\ee}||_{C^{k,a}}$ is uniformly bounded for all small $\ee >0$ by Proposition 2.6
and by \eqref{n1} ,\eqref{n2}. On the other hand  we have the inequality
$$||q_{\ee}||<C\ee ,$$
with the constant $C$ independent of $\ee$. 
Thus we can pass to the limit as $\ee $ goes to $0$ and get
$$\nabla u_{|G_{t_1}} \in C^{k,a}(G_{t_1}).$$

Let $x\in G_t$. Denote by $T$ the tangent plane to $G_t$ at $x$.  Let $\Delta_T $ the Laplace operator on the plane $T$. Since $u\in C^{1,a}(\Omega )$ we can compute $\Delta_Tu(x)$ taking the derivative of $\nabla u$ along $G_t$
and we get for $x\in G_t$,
$$\Delta_Tu(x)\in  C^{k-1,a}(G_{t}).$$

Denote by $M(x)$ the mean curvature of the surface $G_t$ at the point $x$. Let
$T$ be the tangent to $G_t$ plane at $x$, Then
$$\Delta_Tu(x)= M(x) {\pa  u\over \pa  \nu }(x) .$$ 
Therefore 
$$M(x)\in  C^{k-1,a}(G_{t}).$$
Thus if we apply the Schauder estimates for the mean curvature type equation, see [GT],
we get 
$$G_t\in C^{k,a}.$$
The induction step is proved and hence the theorem proved for $x_0\in \Omega $.

Now assume that $x_0\in \pa \Omega$ and the restriction of $u$ on $\pa \Omega$ is in $C^{k,a}$. In this case we need to introduce in \eqref{7}  a correction 
term. Denote by $g$ the restriction on  $\pa \Omega$ of the following function:
$$\int_0^{\infty }f(t)l[G_t(\pa u/\pa \nu )^{-1}]dt.$$

Let $w$ be a solution of the Dirichlet problem
\begin{equation}\label{8}\left\{
\begin{array}{l l}
\Delta w =0, &\mbox{in $\Omega $} \\
w=g &\mbox{on  $\pa \Omega $} \\
\end{array} \right. \end{equation}

Then 
$$u=w+\int_0^{\infty }f(t)l[G_t(\pa u/\pa \nu )^{-1}]dt.$$

Since by Schauder estimates $w\in C^{k,a}(\bar \Omega )$, we can do the estimates
of $u$ as above and we get the desirable estimates near the boundary point. The theorem is proved.

{\bf Proof of Corollary 2.2.} Let $0\in \Omega$. Choose an orthonormal coordinate system
$x_1,...,x_n$ such that $x_1$ has a normal direction to the level surface of $u$ at $0$. 
By Theorem 2.1 all second derivatives of $u$ at $0$ except $\pa^2u/\pa x_1^2$ are bounded
and continuously dependent on point $0$ on the level set.  Since the derivative $\pa^2u/\pa x_1^2$ is uniquely determined by the equation \eqref{6}  and the rest second derivatives are in $C^2$ ($L_{\infty }$), the
theorem follows.

Before going to the proof of Theorem 2.2 we prove some estimates for solutions 
of complex wave equation.

{\bf Proposition 2.6}{\it Let $u\in C^2, \, u=u^1+iu^2,$ be a complex valued solution of an equation

$$ \Box u(t,x) + g^{00}(t,x)u_{tt}+g^{10}(t,x)u_{xt}+g^{11}(t,x)u_{xx}=f $$
$x\in \R, \, 0\leq t \leq T,\,  g$ be a complex valued $C^1$-function  and assume that $u=0$ for large $x$. If

$$|g|=\sum |g^{ij}|<1/2, $$
it follows that 
\begin{equation}\label{9} ||u'(T,\cdot )||\leq 2(||u'(0,\cdot )||+\int_0^T||f(t,\cdot )||dt)exp(\int_0^T2|g'(t)|dt),
\end{equation}
where $||\cdot ||$Êare the $L_2$ norms with respect to $x$ $u'$ is the gradient of $u$ with respect to $x$ and $t$ and 
$$|g'(t)| = \sum sup (|g^{ij}_x(t,\cdot )|+ |g^{ij}_t(t,\cdot )|).$$ }

Proposition 2.6 is Proposition 6.3.2 from [H2] written for complex valued function. It easily
follows from the integration of the identity

$$ 2\Re \bar u_t\Box u = |u_t|^2_t + |u_x|^2_t - 2\Re (\bar u_tu_x)_x, $$
see the proof of Proposition 6.3.2 in [H2].

{\bf Proposition 2.7} {\it Let $u\in C^2, \, u=u^1+iu^2,$ be a complex valued solution be a solution of the quasilinear wave equation 

\begin{equation}\label{10}  \Box u(t,x) + g^{00}(u')u_{tt}+g^{10}(u')u_{xt}+g^{11}(u')u_{xx}=f(x) 
\end{equation}
$x\in \R, \, 0\leq t \leq T,\,  g, f$ be a complex valued $C^2$-function  and assume that $u=0$ for large $x$. Assume that $||u^1(0,\cdot )||_{C^3}<C\, ||u^1_t(0,\cdot )||_{C^2}<C, \, u^2(0,\cdot )=0$,
$||u^2_t(0,\cdot )||_{C^2}<C$,
$\sum |g^{ij}(0)|<1/4$.
Then there exists a constant $T_0>0, T_0<T$ depending on $g, f$ and $C$ such that
for any $0<t<T_0$ $|u''(t,\cdot )|< 2C$. }

{\bf Proof.} There exists a constant $\delta >0$ depending on $g$ such that if
$|u'(t,x)-u'(0)| <\delta $ then $\sum |g^{ij}(t,x)| <1/2$. Applying inequality \eqref{9}  to the second derivatives of equation \eqref{10}  we get
$$ ||u'''(T,\cdot )||\leq 2(||u'''(0,\cdot )||$$
$$+\int_0^T(|g'(t)| ||u'''||+|g''(t)| ||u''||  +||f''(t,\cdot )||)dt)exp(\int_0^T2|g'(t)|dt),
$$
provided that $\sum |g^{ij}(t,x)| <1/2$. Since $||u''||_C\leq ||u'''||$ it follows from Gronwall's lemma that there exists a constant
$T_1$ such that $|u''(t,\cdot )|< 2C$ for $0<t<T_1$ .Set $T_0=\min \{ \delta /2C, T_1\}$.
Then for $0<t<T_0$ $\sum |g^{ij}(t,x)| <1/2$. Thus for $0<t<T_0$ the proposition holds.

Consider the Cauchy problem
\begin{equation}\label{11}\left\{
\begin{array}{l l} \Box u(t,x) =p(u), \\
u(0,x)=u_0,\, u_t(0,x)=u_1
\end{array}\right .\end{equation}
where $-1<x<1,\, 0<t<1,\, u\in \C ,\, u_0\in C^2,\, u_1\in C^1, p$ be a polynomial $p\in \C$.
By [HKM] Cauchy problem \eqref{11}  locally has a classical solution. 

Let $K\ss \R^Ž$ be a triangle with the vertices $(-1,0),\, (1,0),\, (0, 1-\delta )$, where $\delta >0$.

{\bf Proposition 2.8} {\it Let $u$ be a classical solution of the Cauchy problem \eqref{11}  defined in 
 $K$. Assume  $u, \, u''$ are uniformly bounded in $K$.
 Then solution $u$ can be extended as a classical solution of the equation \eqref{11}  in a neighborhood of the point $(0, 1-\delta )$.}
 
 {\bf Proof. } By Proposition 2.7 $u''$ are uniformly bounded in $K$. Taking as initial
 data $u(1-\delta -\epsilon ), \, u_t(1-\delta -\epsilon )$, $\epsilon >0$ be sufficiently small, then
 by the result of [HKM] we get the existence of the solution of Cauchy problem for
 $1-\delta -\epsilon <t<1-\delta +\epsilon $.
 
 As a consequence of the last proposition we have.
 
 {\bf Proposition 2.9}{\it Let $G\ss \R^2$ be a bounded domain with $C^1$ boundary. Assume
$u\in C^3(G)\cap L_{\infty} (\bar G)$ is a solution of the wave equation \eqref{11}  in $G$. Let $z\in \pa G$ and $\pa G$ is not characteristic at $z$. Then $u$ has an extension in a neighborhood of $z$ as a classical solution of the equation \eqref{11} .}

{\bf Proposition 2.10} {\it  Let $u(x,y)$ be a solution of the equation \eqref{6}  in $\Omega \subset \R^2$.
Assume that function $f$ is a polynomial.  Let $G\subset \C^2$ be a bounded domain with a smooth boundary.
Assume that $u$ has a holomorphic extension  on $G$  as a function of complex variables 
$ z_1=x+ix',\, z_2=y+iy'$ and $D^2u$ is bounded in $\bar G$.
Let $z'\in \pa G$ and $T$ be the tangent plane at $z'$. Denote by $L_1$ 2-dimensional plane
$\{ x,y'\}$ and by $L_2$ plane $\{ x',y\}$. Let $l_1,l'_1\subset L_1$  be the lines $x=y'$ and
$x=-y'$.  Let $l_2,l'_2\subset L_2$  be the lines $x'=y$ and
$x'=-y$. Assume that either $T\cap L_1$ is not $l_1, l'_1$, or  $T\cap L_2$ is not $l_2, l'_2$.
Then $u$ has a holomorphic extension in a neighborhood of $z'$.}

{\bf Proof. } Assume that  $T\cap L_1$ is not $l_1, l'_1$. Function $u$ restricted on the planes parallel to $L_1$ satisfies nonlinear hyperbolic equation \eqref{11} . Let point $z\in \pa G$
be sufficiently close to $z'$. Denote by $L_z $ plane parallel to $L_1$, $z\in L_z$.
Then by Proposition 2.10 solution of equation \eqref{11}  can be defined in a neighborhood of $z$.
Considering solutions of \eqref{11}  for points $z\in \pa G$ in a neighborhood  of $z'$ we get
an extension of the function $u$ in a neighborhood  of $z'$ in $\C^2$.
Define smooth in $G$ functions $\psi_j,\, j=1,2, $ by 
$$\psi_j ={\pa y\over \pa \bar z_j}.$$
Taking the derivative of the equation \eqref{10}  with respect to $\bar z_j$ we get that the functions 
$\psi_j$ are solutions of the hyperbolic equations  on $L_z\cap 
G$ . Since $\psi_j$ vanishes on $G$ then by the uniqueness of the solution
of the Cauchy problem it follows that $\psi^j =0$ in $ G$. Thus $u$ satisfies the
Cauchy-Riemann equations in $ G$ and hence $y$ is a holomorphic function in $G$.

{\bf Proof of Theorem 2.2.} We assume first that $f$ is a polynomial. For the polynomial $f$ by  the classical results of S. Bernshtein, [B], and H. Levy, [Le], the solution $u$ is a real analytic function, having holomorphic extension in a domain $G,\, \Omega \subset G \subset \C^2$. Indeed the domain $G$ depends on $f$. The complexification of real variable $x,y$ we will denote by the same letters. Thus we will consider $u(x,y)$ as a holomorphic function of $x,y\in \C^2$. 

We are going to prove that
in the case of a polynomial $f$ the radius of analyticity of the curve $\Gamma $ and the estimates of the complex analytic extension depends only on the third derivative of $f$. Hence the proof
of the theorem will follow after  suitable approximations  of $f$ by polynomials in $C^3$-norm. 

For the proof we consider the complesification of $u$, that allows us to regard $u$ as a solution
of nonlinear wave equation. This method was developed by H. Levy, [Le], and I. Petrovsky, [P].

Choose orthonormal coordinates $x,y$ in $\R^2$ such that $x_0=\{0\}$ and coordinate
$y$ directed along $\nabla u(x_0)$. Then in a neighborhood of $\{0\}$ we can represent the graph
of $u$ as a function $y(x,u)$. Then $y\in C^{3,a} $ in a neighborhood $G$ of $\{0\}$. Clearly the equation \eqref{6}  for the function $y$ takes the form
of a quasilinear elliptic equation,
$$L(y)=\sum a_{ij}(y')y_{ij}-f(u)=0,$$
where $y'$ be the derivatives of $y$.

One can calculate the operator $L$ directly,

\begin{equation}\label{12}L(y)= {y_u^2y_{xx}-y_xy_uy_{xu}+y_x^2y_{uu} \over (y_x^2+y_u^2)^{3/2}(1+(y_x/y_u)^2)}
-{y_{uu}\over  y_u^3}-f(u)=0.\end{equation}

In a neighborhood of zero $y(x,u)$ is a holomorphic function of $x,u\in \C $. If $x_0,\, u_0\in \C$ and $y_u(x_0,u_0)\neq 0$ then function $u$ is holomorphic in a neighborhood of $(x_0,y(x_0,u_0))$.
Denote by $X$ a holomorphic map $X(x,u) =(x,y(x,u))$.
Assume $y$ is holomorphic in a domain $G'\ss \C^2$. Then $y$ satisfies \eqref{12} in $G'$.

Let $z\in G', e\in \C^2$. Denote by $e(z)$ the maximal interval $z+te,\, 0<t<T$ such that
the function $u$ has a holomorphic extension on $e(z)$.

We define a complexification of  solutions $y$. Let $D_R\in \R^2$ be the disk, $|z|<R$. Assume that function $y$ is defined on the disk $D_{2R}$ on the real plane.  Let $r\in \R^2$. Denote, $P(0)\in  \C^2$  the plane  $((0,ir_1),( r_2,0))$,
if $z\ \C^2$ by $P(z)\in  \C^2$ denote the plane $P(0) + z$.

The  equation \eqref{12}  on  $P(z)$  where $y$ is defined has the form
\begin{equation}\label{13}\Box y(r)+ g^{kl}(y')y_{kl}(r)=f(r_2+z).  \end{equation}
where $z$ is a parameter.

Define a 3-dimensional set $H\ss \C^2$. Let $q\in D_{\epsilon }$, $-R<a<R$, 
$H= \{ ((q_1,0),(a,q_2))\}$.  We choose $\epsilon >0$ such small that $y$ is holomorphic on $H$ and for any $q\in D_{\epsilon }$ and $z=((q_1),(iq_2))$, $||f||_{C^2(D_R+z)}\leq  2||f||_{C^2(D_R)}  $. 
Set $e=(i,0)\in \C^2$. Define
$$Z=\cup_{z\in H} e(z).$$
Function $y$ is defined and satisfies equation \eqref{12}  on $Z$. Define function $h$ on $H$
setting $h=(\epsilon^2-q_1^2-q_2^2)(R^2-a^2)$. Denote
$$U_t=\cup_{z\in H} th(z)e,$$
where $t>0$. For sufficiently small $t >0$ we have $U_t\ss Z$.

 From Cauchy-Riemann equations we have
$$y_{z_2}(\cdot ,0)= iy_{z_1}(\cdot ,0)$$

By scaling we may assume without loss that $y_u(0)=-1$. Then $g^{kl}(y'(0))=0$. Hence, since the restriction of $y$ on the real plane is in $C^3$ it follows from Proposition 2.7 that for any $\delta >0$ there exists $c>0$ such for any $0<t<c$, $-c<r_2<c$,
\begin{equation}\label{14}|y'(0, \cdot )-y'(t,\cdot)|<\delta , \end{equation}
 provided by
$|g|^{kl}y'|<1/4$ on $(-c,c)\times (0,c)$ and $y \in Z$. Since a small $\delta >0$ implies the last inequality
it follows the existence of sufficiently small $c>0$ such that inequality \eqref{14}  holds. Constant $c$ depends on
$C^3$-norm of $y$ and on $C^2$-norm of $f$ on the real segment $[u_0-1, u_0+1]$ and is independent on the norms of $f$ in $\C $. From \eqref{14} and Proposition 2.7 it follows that
for  $0<t<c$, $-c<r_2<c$, $|y''|<C$.

We are going to show that $y(t,\cdot )$ is defined for $0<t<c$. Assume not.
Let $t_0$ be the maximal $t$ for which $G_t\ss Z$. 
Let $z_0\in \pa Z\cap \pa U_{t_0}$ 

If $c>0$ is sufficiently small we have $\Re y_{u_1}(z_0) < -1/2$ and hence if $\tilde y$ be the  
restriction of $y$ on $\pa U_{t_0}$ then $\Re \tilde y_{u_1}(z_0) < -1/4$.
Hence $X(\pa U_{t_0})$ is a smooth surface in a neighborhood of $X(z_0)$.
Choosing constant $c>0$ sufficiently small we have the inequality $|\Im \tilde y_{u_1}(z_0)| < 1/8$ and hence the tangent plane at $X(z_0)$ to $X(\Gamma_{s_0})$ satisfies assumptions of Proposition 2.10. Thus by Proposition 2.10 function $u$ has a holomorphic extension in a neighborhood of the point $X(z_0)$ and since $u_y(X(z_0))\neq 0$
function $y$ has a holomorphic extension in a neighborhood of $z_0$.

Thus we proved that  function $y(x,u)$ has a holomorphic extension in the disk
$(x_1+ix_2, 0)$, $x\in D_c$ and bounded in this disk by a constant depending only
on $C^2$-norm of the function $f$. Thus the theorem is proved.

\smallskip

{\bf Proof of Theorem 2.4.} Let $0\in \gamma$. Choose orthonormal coordinates $x,y$ in $\R^2$ such that   coordinate
$y$ directed along $\nabla u(0)$. Then in a neighborhood of $\{0\}$ we can represent the graphs
of $u_i$ as a function $y^i(x,u)$ which satisfy the elliptic equation \eqref{6} . Let $\Gamma $ be the curve
$(y^1, u_1(x,y))$, where $(x,y)\in \gamma $. Set $y=y^1-y^2$. Then $y$ is a solution of
a linear elliptic equation of the form 
 \begin{equation}\label{15}\sum a_{ij}(x)y_{ij} =0, \end{equation}
where $a_{ij}\in C^1$. Uniqueness of the Cauchy problem for the equation \eqref{15}  is well known, see e.g., [H2],  and
since $y=\nabla y =0$ on $\Gamma $ it follows that $y\equiv 0$. Theorem 2.3 is proved. 

\newpage

\section{ Geometry of streamlines }

In this section we prove Theorem 1.3.

{\bf Lemma 3.1} {\it Let $f,y,z: \R\ra\R$, $f(0)=y(0)=z(0)=0$. Let $y\in C^{k,a},\, z\in C^{n-1,a},
\, k,n\in \N, \, a>0$ and $y'(0)>0$.
Assume that $f,y,z$ satisfy the functional equation
\begin{equation}\label{23}z(x)=f(y(x)).\end{equation}
Then $f\in C^{m,a}$, where $m =\min (k, n-1)$.}

{\bf Proof.} Since $y'(0)>0$ it follows that $y^{(-1)}\in C^{k,a}$. Then
$z(y^{(-1)}(x))=f(y(y^{(-1)}(x)))=f(x)$ and the lemma follows. 

Theorem 1.1 follows from Theorem 2.3 and Lemma 3.1. Remarks after Theorem 1.1
follows from Theorem 2.1. Theorem 1.2 is a consequence of Theorem 2.4. 

Let $v$ be a steady flow in $\Omega $ and $v(0)\neq 0$.
The equation \eqref{4}  implies that  that the stream function $u$ of $v$ satisfies the equation
\begin{equation}\label{24}\Delta u = f(u)\end{equation}
in a neighborhood of $0$. If $v(0)=0$ the equation \eqref{24}  might be not satisfied. However,
for an isolated critical point velocity function $v$ satisfies an elliptic equation.

{\bf Proposition 3.2.} {\it  Let $v\in C^2(\Omega )$ be a steady flow and $0\in \Omega $ be an isolated
critical point of $v$, $v(0)=0$. Then in a neighborhood of $0$ $v$ satisfies the equation
\begin{equation}\label{25}\Delta v = c(x)v \end{equation}
where $c\in L_{\infty }$ with the norm depending on $C^2$-norm of $v$. }

{\bf Proof.} By our assumption there is a disk $D$, $0\in D$ such that $0$ is a single critical point 
of the stream function $u$ in $D$. Hence it follows, that for any $x_0\in D$ the connected component $l$ of the
level curve $\{ u(x)=u(x_0) \}$ which contains the point $x_0$ has limit points on $\pa D$. 
Let $x_1\in l\cap \pa D$. Then $\Delta u_i(x_0)=\Delta u_i(x_1)$. By  Lemma 3.1 $|\Delta u_i(x_0)|
\leq C|u_i|$. Therefore the proposition follows.

{\bf Proof of Theorem 1.3.} Let $u$ be the stream function of the flow $v$. 

 If $0$ is a Morse's critical
point of $u$, i.e., Hessian of $u$ at $0$ is not degenerate, then the singularity of $v$ at $0$
is obviously of the type (i). 

Assume now that $D^2u(0)=0$. In a neighborhood of any noncritical point of the stream function $u$  it satisfy the equation $\Delta u=f(u)$ such that $f_u(u(x))= \nabla \Delta u(x)/\nabla u(x)=c(x)$,
where $c$ is a coefficient of equation \eqref{25} . By our assumption $\Delta u(0)=0$, and hence if $l$ is a level set $u(x)=u(0)$ then $\Delta u_{|l}=0$. Set $w=u(x)-u(0)$. Since $c$ is uniformly
bounded we obtain that
\begin{equation}\label{26}\Delta w =d(x)w,\end{equation}
  where $d$ is a bounded function in a neighborhood of $0$.

By
the result of [HO] from equation \eqref{26}  follows that
$$w=p_k +o(|x|^k),$$
where $p_k$ is a homogeneous harmonic polynomial of oder $k$,   $k>2$.
Hence from the equation \eqref{26}  we get
$$\nabla w=\nabla p_k +o(|x|^{k-1}).$$
Therefore,  the singularity
of $v$ is of type (ii).  

Assume finally  that $D^2u(0)$ is degenerate and not equal to zero. Then after rotation axises
the quadratic part of $u$ will be $ax_2^2$. Applying Proposition 3.2 to the first derivative of $u$
we get that the singularity of $v$ at $0$ is of the type (iii).

Consider now the case of stagnating domain $G\ss \Omega$. Since the flow is area preserving
any streamline in $\Omega \sm G $ is either closed or has limit points on $\pa \Omega $.
Denote by $Q$ the union of non-closed  streamlines. If $Q$ has no limit points on $\pa G$ then the theorem follows. If $Q$ has a limit point on $\pa G$ then by topological reason there are no closed streamlines enclosing $G$, and hence $\pa G\ss \pa Q$. Since each streamline from $Q$ has limit points on  $\pa \Omega $ it follows as in the proof of Proposition 3.2 that on $Q$ $v$
satisfies equation \eqref{25} . Since $v\equiv 0$ on $G$ and $c\in L_{\infty }$ then $v\equiv 0$  in 
$\Omega $ by the unique continuation theorem if $\pa G$ is the limit set for $Q$.
The theorem is proved.

\section{Stable instability of unstable Arnold flows }

In this section we prove Theorem 1.5. The main idea of the proof is similar to the approach  we used for
the proof of the existence of the wandering trajectories to the solution of the Euler equation, [N].

Let $r, \theta $ be the polar coordinates in $A$. Let $\tilde v\in C^{2,a}(A)$ be a radial symmetric Arnold stable steady flow, $\tilde v=\tilde v(r)$. Since the flow $\tilde v$  satisfies inequalities \eqref{5}  it follows that $\tilde \omega_r > 0$ and we may assume without loss that
$\tilde \omega_r>1$ in $A$. Hence, if $\tilde v^1,\tilde v^2$ be the components of $\tilde v$ in the coordinates $r, \theta $ 
then

$$\pa \tilde v^2/ \pa r>1 .$$

  \begin{equation}\label{27} \pa \tilde \omega / \pa r>1.  \end{equation}

 Let $h\in C^1(A)$ and $\pa h/ \pa r > 0$.
 
 Set
 $$h^+ = \sup_{x\in A}{ \pa h(x)/ \pa \theta \over |\nabla h(x)|} ,$$
 $$h^- = \inf_{x\in A}{ \pa h(x)/ \pa \theta \over |\nabla h(x)|} ,$$
 $$h^*=h^+-h^-.$$
 Thus if $h^*=0$ then $h$ is a function of radius. 
 
 \smallskip
 
 {\bf Lemma 4.1.}{\it There is a $\delta >0$ such that if $||v(x,t)-\tilde v(x)||_{C^{2,a}(A)}<\delta $ then the inequality $\omega^+>-\omega^-$ yields, 
  $$ {\pa \omega^+ \over \pa t}(0)>c_0\omega^+,$$
and the inequality $\omega^+<-\omega^-$ yields, 
 $$ {\pa \omega^- \over \pa t}(0)>-c_0\omega^-,$$ }
 where $c_0$ be a positive constant
 
 {\bf Proof.} Let $g^t$ be a one parametric group of diffeomorphism of $A$ corresponding to the flow $v(t)$. Then from Euler-Helmholtz equation follows, see [AK],
 $$\omega (x,0)=\omega (g^t,t).$$  
 Thus 
  \begin{equation} \label{28}{\pa \nabla \omega (g^t(x),t) \over \pa t }(0) =J_v^T\nabla \omega (x,t), \end{equation}
 where $J_v$ is the Jacobian matrix of the vector field $v(x,0)$. Notice that

 $$J_{\tilde v}=      \left(%
\begin{array}{cc}0
    & 0  \\
   a&  0
\\
\end{array}%
\right),     $$
where $a>1$.
 
 Let $u,\, \tilde u$ be the stream functions of $v(\cdot , 0)$ and $\tilde v$. Denote
$u'=u- \tilde u$, $\omega' =\omega (\cdot, 0)-\tilde \omega $. Then $u'$ satisfies the equation,
$$\Delta u'= \omega' . $$
On the boundary circles  of $A$ $u'$ equal to constants bounded by $\delta $. From the standard estimates for the solutions of the Poisson equation it follows, that $|u'_{12}|<K\delta $ or

$$ |v^1_1|<K\delta ,  $$
 where $K$ is a positive constant.  For $u'_{\theta }$ we have the 
equation
$$\Delta u'_{\theta }= \omega'_{\theta }\,\,\, in \,\,\, A , $$
$$u'=0 \,\,\, on \,\,\, \pa A.$$
Then
$(u'_{\theta })_{|C^1(A)} \leq K(\omega'_{\theta })_{|C(A)}$. Since 
$|\omega'_{\theta } | < \delta \omega ^++\omega^-$ we get
$$|\nabla (v^2-\tilde v^2)| < \delta \omega ^++\omega^-.$$
Thus for sufficiently small $\delta >0$ we have

  \begin{equation}\label{29}|v^2_1|>1,\,\, |v^2_2|<K\delta \omega^++\omega^-,\,\, |v^1_2|<K\delta \omega^++\omega^-,\,\, |v^1_1|<K\delta ,  \end{equation}
  
  Let    $x_0,x_1\in A$ and
   $${ \pa \omega (x_0,0)/ \pa \theta \over |\nabla \omega (x_0,0)|}=\omega^+,\,\, \,\, { \pa \omega (x_1,0)/ \pa \theta \over |\nabla \omega (x_1,0)|}=\omega^- .$$.
  
   Assume first that $\omega^+\geq -\omega^-$. Then  from \eqref{29}  we get
  \begin{equation}\label{30} \omega_1 (x_0,0)>c,\,\,\, | \omega_2 (x_0,0)|< C\omega^+.   \end{equation}
  
  Denote
  $$(a,b)={\pa \nabla \omega (g^t(x),t) \over \pa t }(0).$$
  Then from \eqref{28} , \eqref{29} , \eqref{30}  we get $a>c-C\delta \omega^+,\, |b|<C\delta \omega^+$, where
  $C$ is a positive constant. Since
  $$ {\omega_2 (x_0,0)\over \omega_1 (x_0,0)}=\omega^+ $$
  we get that $\pa \omega^+(0)/\pa t>c_0\omega^+$.
  
  If we assume now that $\omega^+\leq -\omega^-$. Then after the similar computations at
  the point $x_1$ we get that  $\pa \omega^-(0)/\pa t> -c_0\omega^-$. The lemma is proved. 
  
  {\bf Proof of Theorem 1.5.} Assume by contradiction  that for all $t>0$ $||v(0,t)-\tilde v(x)||_{C^{2,a}(A)}<\delta $.

Assume that $\omega^+(t_0)=-\omega^-(t_0)$, $t_0\in  \R$. Then by Lemma 4.1 $\pa \omega^+(t_0)/\pa t >0,\,\, \pa \omega^-(t_0)/\pa t >0 $ and hence for all
  $t>t_0$ $\omega^+(t)>-\omega^-(t)$. Hence for $t>t_0$ we have  $\pa \omega^+/\pa t>c_0\omega^+$. Then there is $T>t_0$ such that for $t>T$ $||v(0,t)-\tilde v(x)||_{C^{2,a}(A)}>\delta $.
  
  Assume now that for all $t>0$, $\omega^+(t)<-\omega^-(t)$.  Then we have  $\pa \omega^-/\pa t<-c_0\omega^-$ and therefore $\omega^+(t), \omega^-(t)$ tend to $0$ as $t\ra \infty$.  Hence 
  $\omega (\cdot,t) \ra \bar \omega $, where $\bar \omega$ depends only on $r$. Let $\bar v$
  be the velocity corresponding to the vorticity $\bar \omega$. Then $\bar v$ is a steady flow 
  corresponding to the vorticity $\bar \omega $. Since by our assumption $||\bar v-\tilde v(x)||_{C^{2,a}(A)}<\delta $ the flow $\bar v $ is Arnold stable and hence by the theorem of Arnold,  [A],
  the trajectory $v(\cdot, t)$ can not tend to $\bar v$. Thus we got a contradiction. Theorem 1.5 is
  proved.

\section{ Variational solutions of Euler equation }

In this section we study Arnold stable steady state flows and prove Theorem 1.4 and 1.6.

{\bf Theirem 5.1.}{\it Let $v\in C^2(\Omega )$ be an Arnold stable solution of \eqref{2} , \eqref{3} . Then $\omega $ is of a constant
sign. In any compact subdomain of $\Omega $ the critical set of $v$ is in a union of a finite collection
of $C^2$ curves. If $\Omega $ is a convex domain then the critical set of $v$ is a single point.}

{\bf Proof. } Let $u$ be a stream function of $v$, $u=0$ on $\pa \Omega$. Denote by $\Sigma $
the set of critical points of $v$, and by $\Sigma_0$ the interior of $\Sigma$.

On the set $\Omega \sm \Sigma $ $v$ is a solution of equation \eqref{25}, where 
$$c(x)=\nabla  \Delta u / \nabla u ,$$
$c\in L_{\infty }, \, c>0$. Since $v\in C^2$ one can define equation \eqref{25} on $\Sigma \sm \Sigma_0$
by continuity with $c\in L_{\infty }$.

If $\Sigma_0$ is nonempty  then, since $\Sigma_0$ is an open set and $v=0$ on $\Sigma_0$
we have by unique continuation theorem, see [H1], $v\equiv 0$. Hence $\Sigma_0$ is empty. 
Therefore, $v$ satisfies equation \eqref{25}  in the whole domain $\Omega$. The critical set 
of $v$ is the intersection of the nodal lines of $v^1$ and $v^2$. Since solutions of \eqref{25}
have isolated second order zeros, see [HO], the second part of the theorem follows.

From the boundary condition \eqref{2} it follows that $\omega =m=const$ on $\pa \Omega$.
We show that $m\neq 0$. Assume by contradiction that $m=0$. Denote by $N$ the nodal set
of the function $u$. We show first that there is a nodal domain $G\ss \Omega\sm N$ which has
limit points on $\pa \Omega$.  If there is no such domain it implies that any curve in $\Omega$ with the end point on $\pa \Omega $ has infinitely many intersection with $N$. Hence $u$ has 3-d order zero on $\pa \Omega$. Thus $v$ has the second order zero on $\pa \Omega$ and by the uniqueness of the Cauchy problem for equation \eqref{25}, [H1], it follows that $v\equiv 0$.

Thus there exists a nodal domain of $u$ $G$ which has limit points on $\pa \Omega$. Thus the
vorticity $\omega $ vanishes on the exterior component of the boundary  of $G$. Assume without loss
that $u>0$ in $G$. From the structure
of the critical points of $v$ in $\Omega$ easily follows the existence of a rectifiable curve
$\gamma :[0,1]\ra G$ such that $(\dot \gamma ,\nabla u) \geq 0$,  $\gamma (0)\in \pa \Omega$
and $\gamma (1)$ is a local supremum of $u$. From the inequality \eqref{5} follows that 
$\omega (\gamma (1))>0$. Since $\gamma (1)$ is a local supremum of $u$ the last inequality
contradicts the maximum principle.

Thus we prove that $m\neq 0$. We assume without loss that $m<0$. We prove that
$\omega <0$ in $\Omega$.
Assume by contradiction that there is a nodal domain $D$ of  $\omega $ where $ \omega(x)>0$. Assume that $G$ is non empty domain. Let at a point $z\in \bar D$
$u$ attains its supremum over $\bar D$. From inequality \eqref{5}  it follows, that $\omega (z)>0$.
Thus $z\in D$. By the maximum principle point $z$ can not be a point of a local supremum of $u$. Thus it follows that $D$ is an empty set, and hence the first part of the  theorem is proved.

Assume now that $\Omega$ is a convex domain. We prove that then $v$ has a single critical point. We will use some arguments suggested in [CC]. Let $e\in \R^2, \, |e|=1$. Denote by $e_1, e_2$ two points on $\pa \Omega$ where $e$ is tangent to the boundary. Consider the derivative
of stream function $u_e$.  Then $u_e$ is a solution of equation \eqref{25}. Denote by $\gamma_e$ the nodal line of $u_e$. Since $c\geq 0$ the maximum principle holds for the solutions of 
\eqref{25}. Therefore $\gamma_e$ can not enclosed  any subdomain in $\Omega $. Then it
follows that $\gamma_e$ is a simple arc with the end points $e_1,e_2$. Let $z_1,z_2\in \Omega$
be two different critical points of $v$. Then $z_1,z_2\in \gamma_e$ for all $e\in S^1$. Notice
that $\gamma_e$ continuously depends on $e\in S^1$. Thus oder of the points $e_1,z_1,z_2,e_2$ along the arc $\gamma_e$ is independent on $e$. On the other hand
when $e$ is changing to $-e$ the curve $\gamma_e$ changes it orientation. That leads to
a contradiction with the existence of the second critical point of $v$. The theorem is proved.

From Theorem 5.1  follows Theorem 1.5.  

{\bf Lemma 5.2.} {\it Let $v$ be a smooth steady flow. Assume that in a neighborhood of $0$,
\begin{equation}\label{31}\omega =c+l_1l_2l_3+o(|x|^3),\end{equation}
where $l_i$ are three linear functions,  each two are linear independent. Then $c=0$.}

{\bf Proof. } Let $u$ be a stream function of $v$. Vorticity $\omega $ is a constant on connected
components of the level sets of $u$. By our assumption that is impossible in a neighborhood of $0$ 
if $Du(0)\neq 0$ or $D^2u(0)\neq 0$. Thus $D^2u(0)=0$ and hence $c=0$. The lemma is proved.

{\bf Proof of Theorem 1.6.} By theorem  of Burton, [B2], there exists a steady state solution
$v$ with $\omega \in \bar S(\omega )$ with the stream function $u$ such that
 $\Delta u =f(u)$,
with $f\in L_{\infty }$ being a monotonically non-decreasing function. Thus
 $u\in C^{1,a}$, $a>0$. By Theorem 2.1 and Theorem 1.4 the level
lines of $u$ are smooth curves. Let $m_1, m_2$ be functions of distribution of $\omega $ and $u$,
$$m_1(t)=|\omega^{-1}(0,t)|,\;\; m_2(t) =|u^{-1}(0,t )|.$$ 
Set $z= m_1^{-1},\, y=m_2^{-1}$. Functions $z,y$ are defined on $(0,m)$, where $m=|\Omega |$. 
Function $z$ is smooth on $(0,1)$ by the assumptions, $y\in C^{1,a}$ since $\pa u /\pa \nu$
does not vanish on the level curves of $u$. Since of functional equation $z(t)=f(y(t))$ we get
by Lemma 3.1 $f\in C^{1,a}  $.  Hence from equation \eqref{6}  we get $u\in C^{3,a}$
in $\Omega \sm s$, where $s$ is a point of supremum of $u$. Iterating the argument
we get the smoothness of $u$ in  $\Omega \sm s$. The theorem is proved.

\smallskip

{\bf Remark.} Let $h$ be a smooth positive function in $\Omega $, $h=0$ on $\pa \Omega $ and
$h$ has a singularity of type \eqref{31}  at $0\in \Omega$. Assume that variational problem (A) has a 
smooth extremal $\omega = h(g)$, where  $g\in S\, Diff\, (\Omega )$. Then $\omega $ has a
singularity of the type \eqref{31}  at $g(0)$. Hence, by Lemma 5.2 $\omega (g(0))=0$. That contradicts
the assumption on the positivity   of $h$ and therefore the variational problem (A) has no smooth
extremals.  

\smallskip

{\bf Proposition 5.4. }{\it Let $\Omega \ss \R^2$ be a bounded domain with a smooth boundatry. Let $u_1,u_2$ solutions of the Dirichlet problem
\begin{equation}\label{34}\left\{
\begin{array}{l l}
\Delta u_i =f_i(u_i), &\mbox{in $\Omega $} \\
u_i=0 &\mbox{on  $\pa \Omega $} \\
\end{array} \right. \end{equation}
where $f_i$ are negative increasing $C^1$-functions. 
Then $f_1(u_1) \notin \bar S(f_2(u_2))$. }

{\bf Proof.} Assume by contradiction that
$$f_1(u_1) \in \bar S(f_2(u_2))$$
$ S^*(f_2(u_2))$ be a weak closure of $\bar S(f_2(u_2))$. Then $ S^*(f_2(u_2))$ is a convex set and hence if $u^*$ be a global minimizer of the variational problem (K) on $ S^*(f_2(u_2))$ 
then $u^*$ also a global minimizer of (K) on $ S^*(f_2(u_2))$, [B1]. We may assume without loss that
\begin{equation}\label{35}u_2=u^*.\end{equation}
Set $w=u_2-u_1$. Compute variation of the kinetic  energy $E$  along $w$ at $u_1$,
$$\delta E_{|w}(u_1)=\int_{\Omega}\nabla u_1\nabla w dx =-\int_{\Omega}wf_1(u_1)dx .$$
 
Since $f_1<0,\, f'_1>0$ and by \eqref{34}, \eqref{35}  we have
$$\int_{\Omega}u_1f_1(u_1)dx   \geq \int_{\Omega}u_2f_1(u_2)dx   \geq \int_{\Omega}u_2f_1(u_1)dx$$
Hence
$$\delta E_{|w}(u_1)\geq 0.$$
By assumption \eqref{35}
$$\delta E_{|-w}(u_2)= 0.$$
Since the energy $E$ is a convex function on the line connecting points $u_1$ and $u_2$
we get a contradiction. The proposition is proved.

\bigskip
 \centerline{REFERENCES}
 
 \medskip
 \noindent [AM] S. Alinhac, G. M\'etivier
{\it Propagation de l'analyticit\'e locale pour les solutions de lÕ\'equation dÕEuler, }
 Arch. Rational Mech. Anal. 92 (1986), 287Ð296.
 
  \medskip
 \noindent [AP] S. Alpern, V.S. Prasad
{\it Typical dynamics of volume preserving homeomorphisms,} Cambridge University Press, Cambridge, 2000.
 
 \medskip
 \noindent [A1]
 V. Arnold, {\it Sur la g\'eom\'etrie differentielle des groupes de Lie de dimension infinie et ses applications \`a l'hydrodynamique des fluids parfaits,} Ann. Inst. Fourier, 16, (1966),  319Ð 361

\medskip
 \noindent [A2] V.I. Arnold {\it On an apriori estimate in the theory of hydrodynamical stability, }
 Amer. Math. Soc. Transl. 19 (1969), 267-269. 
 
 \medskip
\noindent [AK] V.I. Arnold, B.A. Khesin {\it Topological Methods in Hydrodynamics },
Sringer, 1998.

 \medskip
 \noindent [BBZ] C. Bardos, S. Benachour, M. Zerner
{\it Analyticit\'e des solutions p\'eriodiques de l'\'equation d'Euler en deux dimensions }
C. R. Acad. Sci. Paris S\'er. A-B 282 (1976), no. 17, 995-998.

 \medskip
\noindent [B] S. Bernshtein {\it D\'emonstration du th\'eor\`eme de M. Hilbert sur la nature analytique des solutions des \'equations du type elliptique sans l'emploi des s\'eries normales, }Math. Zeitschrift, 28 (1928),  330Ñ348.

\medskip
 \noindent [B1]
G.R. Burton  {\it Variational problems on classes of rearrangements and multiple configurations
for steady vortices,} Ann. Inst. Henri Poincar\'e 6 (1989), 295-319.

\medskip
 \noindent [B2]
G.R. Burton  {\it Rearrangements of functions, saddle points and uncountable families of steady configurations for a vortex.} Acta Math. 163 (1989), 291Ð309.

\medskip
 \noindent [CC]
X. Cabr\'e, S. Chanillo {\it Stable solutions of semilinear elliptic problems in convex domains}, Selecta Math. (N.S.)  4 (1998), 1-10.

\medskip
 \noindent [C] J-Y. Chemin {\it Perfect Incompressible Fluids }, Clarendon Press. Oxford, 1998

\medskip
 \noindent [EM]
D. Ebin, J. Marsden, {\it Groups of diffeomorphisms and the motion of an incompressible
fluid,} Ann. Math., 92 (1970), pp. 102Ð163.

\medskip
 \noindent [GT] D. Gilbarg, N. Trudinger, {\it Elliptic Partial
Differential Equations of Second Order, 2nd ed.}, Springer-Verlag,
Berlin-Heidelberg-New York-Tokyo, 1983.

\medskip
 \noindent [G] N.M. G\"unter {\it Potential Theory and its Application to Basic Problems of Mathematical Physics }, New York. Frederick Ungar Publishing Co., 1967

\medskip
 \noindent [HNY] F. Hamel, N. Nadirashvili, Y. Sire
{\it On the level sets of solutions of elliptic problems in convex domains or convex rings in arbitrary dimension}, Preprint

 \medskip
 \noindent [HO] 
M. Hoffmann-Ostenhof, T. Hoffman-Ostenhof {\it Local properties of solutions of Schrodinger
equations, } Comm. PDE 17 (1992), 491Ð522.

 \medskip
 \noindent [HON] 
M. Hoffmann-Ostenhof, T. Hoffman-Ostenhof, N. Nadirashvili {\it Interior H\"older estimates for solutions of Schr\"odinger equations and the regularity of nodal
sets,}
Comm. PDE 20 (1995),  1241-1273.

\medskip
 \noindent [H1]  L. H\"ormander {\it Linear Partial Differential Operators,} Springer, 1963 

\medskip
 \noindent [H2]  L. H\"ormander {\it Lectures on Nonlinear Hyperbolic Differential Equations }
 Springer, 1997.
 
\medskip
 \noindent [HKM] T.J.R. Hughes, T. Kato, J.E. Marsden
{\it Well-posed Quasi-linear Second-order Hyperbolic Systems with Applications to Nonlinear Elastodynamics and General Relativity,} Arch. Rational Mech. Anal. 63 (1976), 273 -294

\medskip
 \noindent [Ka]
L.I. Kamynin {\it The smoothness of heat potentials, part 5,}  Diff Equat., 4, 185-195 (1968), 185-195

 \medskip
 \noindent [K]
H.  Koch {\it Transport and instability for perfect fluids,} Math. Ann. 323 (2002), no. 3, 491Ð523.

\medskip
 \noindent [Le] H.L e w y {\it 	\"Uber 	den	analytischen	Charakter	der	L\"osungen	elliptischer	Diffe- rentialgleichungen,} Nachrichten von der Gesellschaft der Wissenschaften zu G\"ottingen, Math.-Phys. KL, 1927,  178-186.

\medskip
 \noindent [Li]  L.Lichtenstein {\it \"Uber einige Hilfss\"atze der Potentialtheorie },  Math. Zeitschrift
23, 72-78 (1925).

\medskip
 \noindent [L] P-L. Lions  {\it Mathematical Topics in Fluid Mechanics. v.1. Incompressible Models,}
 Claredon Press. 1996

 \medskip
 \noindent [M] J. Milnor {\it Remarks on infinite-dimensional Lie groups,} Proc. Summer School on 
 Quantum Gravity, Amsterdam, 1984, 1007-1057

 \medskip
 \noindent [Mi] C. Miranda {\it Partial Differential Equations of Elliptic Type. 2nd ed.,} Springer,
 1970

\medskip
 \noindent [N] N. Nadirashvili {\it Wandering solutions of the Euler 2D equation,} Funct. Anal. Appl.
 25 (1991), 220-221.
 
 \medskip
 \noindent [P]  I. G. Petrowsky {\it  Sur l'analyticit\'e des solutions des syst\`emes d'\'equations diffŽrentielles,}
Mat. Sbornik, 5(47) (1939),  3-70

 \medskip
 \noindent [S] A. Shnirelman
{\it Evolution of singularities, generalized Liapunov function and generalized integral for an
ideal incompressible fluid,} Amer. J. Math. 119 (1997), 579Ð608.
 
 \medskip
 \noindent [W] M. Wiegner 
{\it Schauder estimates for boundary layer potentials }, Math. Methods Appl. Sci. 16 (1993), 877-
894.

 \medskip
 \noindent [Y] V.I. Yudovich {\it Non-stationary flow of an ideal incompressible liquid,} Zhurn. 
 Vych. Mat. 3 (1963), 1032 - 1066

\end{document}